\documentclass[3p]{elsarticle}

\usepackage{longtable,tabularx}
\usepackage{hyperref}
\usepackage{mathtools}
\usepackage{makecell}
\usepackage{diagbox}
\usepackage{multirow}
\usepackage{siunitx}

\usepackage{amssymb}
\usepackage{amsmath}

\usepackage{lineno}

\journal{Renewable Energy}

\begin{document}

\begin{frontmatter}



    \title{Novel discretization method to calculate $g$-functions of vertical geothermal boreholes with improved accuracy and efficiency} 


    \author[1]{Yue Yang}
    \author[1]{Xiaodong Yang}
    \author[2]{Chenhui Lin}
    \author[3]{Luo Xu}
    \author[4]{Qi Wang}
    \author[2]{Shuwei Xu}
    \author[2]{Wenchuan Wu}

    \affiliation[1]{organization={State Key Laboratory of High-Efficiency and High-Quality Conversion for Electric Power, Hefei University of Technology},
        city={Hefei},
        postcode={230009},
        country={China}}
    \affiliation[2]{organization={State Key Laboratory of Power Systems, Department of Electrical Engineering, Tsinghua University},
        city={Beijing},
        postcode={100084},
        country={China}}
    \affiliation[3]{organization={Department of Civil and Environmental Engineering, Princeton University},
        city={Princeton},
        postcode={08544},
        country={USA}}
    \affiliation[4]{organization={The Hong Kong Polytechnic University},
        city={Hong Kong},
        postcode={999077},
        country={China}}

    \begin{abstract}
        The calculation of $g$-functions is essential for the design and simulation of geothermal boreholes. However, existing methods, such as the stacked finite line source (SFLS) model, face challenges regarding computational efficiency and accuracy, particularly with fine-grained discretization. This paper introduces a novel discretization method to address these limitations. We reformulate the $g$-function calculation under the uniform borehole wall temperature boundary condition as the solution to spatio-temporal integral equations. The SFLS model is identified as a special case using stepwise approximation of the heat extraction rate. Our proposed method employs the Gauss-Legendre quadrature to approximate the spatial integrals with a weighted sum of function values at strategically chosen points. This transforms the time-consuming segment-to-segment integral calculations in SFLS model into simpler and analytical point-to-point response factors. Furthermore, we identify that the governing integral equations are of the Fredholm first kind, leading to ill-conditioned linear systems that can cause $g$-function to diverge at high discretization orders. To address this, a regularization technique is implemented to ensure stable and convergent solutions. Numerical tests demonstrate that the proposed method is significantly more efficient, achieving comparable or improved accuracy at speeds 20 to 200 times faster than the SFLS model with optimized nonuniform discretization schemes.
    \end{abstract}




\end{frontmatter}



\section{Introduction}

Geothermal boreholes are the critical interface for heat exchange in ground source heat pump (GSHP) \cite{sarbu2014general, luo2023overview} and borehole thermal energy storage (BTES) systems \cite{wang2024borehole}. These technologies are cornerstones in the global effort to decarbonize building heating and cooling, leveraging the earth's stable subsurface temperature for exceptional energy efficiency. Coupled with BTES, these systems further enable the seasonal shifting of thermal energy, storing excess heat or cold from renewable sources or waste heat for later use. Such integrated solutions not only reduce reliance on fossil fuels but also enhance grid resilience by balancing supply-demand mismatches in renewable-dominated energy systems.

The thermal performance of these systems hinges on the heat exchange capability of the boreholes, which is influenced by their wall temperatures and the ground's thermal properties. Consequently, the efficiency and economic viability of GSHP and BTES systems depend critically on the accurate modeling and simulation of geothermal boreholes, particularly in predicting their long-term thermal response. It is crucial to simulate the temporal evolution of borehole wall temperature accurately with respect to specific heat load patterns to evaluate the long-term thermal performance of geothermal boreholes from a system design and optimization perspective.

$g$-function, namely the dimensionless thermal response function, is one of the most important tools in the simulation of geothermal boreholes since it is introduced by Eskilson \cite{eskilson1987thermal}. It has been widely used to design and optimize the performance of GSHP \cite{rees2016advances, west2024ground} and BTES systems \cite{liu2021recent}. Once the $g$-function of a group of boreholes is known, the effective borehole wall temperature $T$ can be written as a convolution in time domain:
\begin{equation}
    T(t) = T_g - \frac{1}{2 \pi k} \int_{0}^{t} Q(t-\tau) \frac{dg}{dt}(\tau) \ d\tau
\end{equation}
where $T_g$ is the undisturbed ground temperature, $k$ is the thermal conductivity of the ground, $Q(t)$ is the average heat extraction rate per length of borehole, and $g(t)$ is the $g$-function of borehole field. The $g$-function is a function of time that describes the temperature response of the borehole wall to a unit heat extraction rate per length applied at time $t=0$. The temporal convolution can be computed efficiently with fast fourier transform (FFT) \cite{marcotte2008fast,chen2019thermal} and various load aggregation methods \cite{bernier2004multiple,claesson2012load}.

The computation of $g$-function is a challenging task due to the complex heat transfer processes involved inside and outside geothermal boreholes. Numeric finite difference models can be used to solve the partial differential equations governing thermal process and obtain the values of $g$-functions, including the superposition borehole model (SBM) proposed by Eskilson \cite{eskilson1987thermal}, the duct ground storage model (DST) used in TRNSYS \cite{pahud1997duct} and more general finite element methods (FEM) software like COMSOL \cite{law2016characterization} or OpenGeoSys \cite{yang2024numerical}. However, numerical methods are computationally expensive and time-consuming, especially for large bore fields with many boreholes and long simulation periods, which drives the development of analytical models of $g$-functions for clear physical interpretation and more efficient calculation.

Among analytical models, finite line source (FLS) is the most widely used approach to analyze the thermal behavior of geothermal boreholes, which ignores the geometric details of boreholes and assumes that each borehole is a line heat source with finite length in a semi-infinite ground. The influence of heat exchange on the temperature of borehole walls can be acquired via spatial and temporal superposition. The study of FLS model can be dated back to the original definition of $g$-function by Eskilson \cite{eskilson1987thermal}, then rediscovered by various researchers including Zeng \cite{zeng2002finite}, Lamarche \cite{lamarche2007new} and broadened to inclined boreholes in \cite{cui2006heat} and \cite{marcotte2009effect}.

These early studies on FLS usually assumes that the heat exchange rate is a constant value along the depth of borehole and calculates their influence on the borehole wall temperature at specific points or the average temperature of the whole borehole. This assumption leads to the depth-dependent distribution of borehole wall temperature and is not well aligned with the boundary condition used by Eskilson's original SBM model to compute $g$-function, which assumes that the borehole wall temperature is uniform along the depth of borehole and the heat exchange rate is a depth-dependent variable. The discrepancy between the boundary conditions results in the overestimation of $g$-function given by FLS model compared to SBM model, as pointed out by several researchers in \cite{lamarche2007new,lamarche2009fast,fossa2011temperature}.

The seminal work of Cimmino \cite{cimmino2014semi} fulfills the gap of boundary conditions by dividing the borehole into segments and assuming that the heat exchange rate is constant in each segment but different across segments, which is called stacked finite line source (SFLS) model. They discuss three kinds of boundary conditions (BC) for FLS model of boreholes, where the depth-agnostic constant heat exchange rate is refered as BC-I and the uniform borehole wall temperature (UBWT) is named after BC-III.
The SFLS model can approximate the uniform borehole wall temperature boundary condition (BC-III) used in SBM model by constraining the average temperature of each segment to be equal, and the $g$-function calculated by SFLS model is consistent with SBM model with significantly lower computational time.
The methodology of SFLS model under BC-III is validated by FEM models with similar boundary conditions in \cite{monzo2015novel} and generalized from vertical boreholes to inclined boreholes in \cite{lazzarotto2016methodology,lazzarotto2016methodology2}. Cimmino further extends the SFLS model to calculate $g$-function under more boundary conditions considering the thermal resistance \cite{cimmino2015effects}, the connection topology of boreholes \cite{cimmino2019semi} and variable mass flow rates \cite{cimmino2024g}. Open source implementation of SFLS model is also available in the \texttt{pygfunction} package \cite{cimmino2018pygfunction,cimmino2022pygfunction} with support for various boundary conditions and borehole configurations.

The accuracy and efficiency of SFLS model depends on the number of segments and the length of each segment to discretize the borehole and approximate the depth-varying heat exchange rate. Increasing the number of segments generally leads to a more accurate representation of the borehole's thermal response, as it allows for a finer resolution of the heat exchange process along the borehole's depth. Researchers have observed that the number of segments per borehole required to achieve acceptable precision of $g$-function increases as the number of boreholes grow \cite{cook2021development}. Nevertheless, the computational time of SFLS model increases quadratically with respect to the total number of segments \cite{cook2021development} because the segment-to-segment heat response factor has to be computed for each pair of segments, so the number of segments is usually limited to a small value in practice. Cimmino \cite{cimmino2014semi} tests the accuracy of SFLS model up to 256 segments and suggests to use 12 segments per borehole as an approximation. Lamarche \cite{lamarche2017g} proposes a revised SFLS model to assume linear variation of heat extraction rate within each segment and concludes that 8 segments per borehole is sufficient to achieve accuracy of 12 segments per borehole with conventional SFLS. Cook \cite{cook2021development} explores the idea of nonuniform discretization of boreholes, where the segments are denser near the beginning and ending of borehole to describe the rapid variation of heat extraction rate, and shows that 8 unequal segments can achieve better accuracy than uniform discretization with the same number of segments. Their sequential research \cite{cimmino2024optimal} aims to optimize the nonuniform discretization scheme by minimizing the error of $g$-function tested on a large corpus of 557056 bore field configurations. However, existing work on the discretization scheme of SFLS model is mostly based on empirical experimentation on numerous simulations and lacks formal theoretical foundations. It is still an open problem on the existence of better discretization scheme to reduce both the error and computational time of $g$-function for geothermal boreholes.

In this paper, we aim to provide new insights on the $g$-function of boreholes under UBWT boundary condition from the perspective of integral equations and propose a novel discretization scheme to improve the accuracy and efficiency of $g$-function computation compared with existing SFLS model. The main contributions of our paper can be summarized as answers to the following questions that are crucial for calculating $g$-functions of boreholes:

\begin{enumerate}
    \item \textbf{Why does detailed discretization of boreholes lead to more accurate $g$-function?} \\
          We find out that $g$-function is essentially the solution to a group of spatio-temporal integral equations describing the thermal process of boreholes under specific boundary conditions, and the segmentation of borehole can be regarded as discrete approximation of the original continuous heat extraction rate with respect to the depth and its spatial integral. The error of approximation naturally decreases as the order of discretization grows because it captures more spatial details of the heat exchange process on the borehole wall. Under this framework, the widely-used SFLS model is a special discretization scheme of the integral equation by assuming the heat exchange rate as a stepwise function that remains constant in each segment.

    \item \textbf{How can we acquire more accurate $g$-function with limited order of discretization?} \\
          We propose a novel discretization scheme that approximates the spatial integral of heat extraction with Gauss-Legendre quadrature rule, which can achieve better accuracy than the SFLS model with the same order of discretization due to the rapid convergence of Gauss-Legendre quadrature. Instead of assuming a constant heat exchange rate within each segment, this approach only requires the value of heat extraction rate at some strategically chosen points at specific depths, so the segment-to-segment coupling factors in SFLS that are time-consuming to calculate as integral are reduced to much simpler point-to-point coupling factors with analytical form. Thus, the computation of $g$-function can be significantly accelerated due to the faster convergence rate with respect to discretization and more efficient formulation of the discretized equations. It is observed that the proposed method can obtain the $g$-function with similar or lower error at 20-200 times faster speed compared to SFLS model.

    \item \textbf{Will the value of $g$-function always converge if the order of discretization is sufficiently large?} \\
          It is an intuitive assumption that increasing order of discretization will lead to a converged value of $g$-function. However, this conjecture is not thoroughly discussed or proved in previous research.
          In this paper, we discover that the integral equations under UBWT boundary condition belong to the Fredholm integral equation of the first kind. Its discretized linear forms will be ill-conditioned if the order of discretization becomes too large and may fail to provide a meaningful value of $g$-function. The divergence of $g$-function is demonstrated by our numerical tests for both SFLS model and the proposed discretization method. A regularization technique is developed to overcome the ill-conditioning and ensure the stable convergence of $g$-function under arbitrary discretization. As far as we know, this is the first time that the convergence of $g$-function is discussed in the context of integral equations and the unintuitive divergence of $g$-function is explained by the ill-condition of discretized equations.
\end{enumerate}

The paper is structured as follows. In Section \ref{section: methodology}, we present the methodology used in this study, including the formulation of the integral equations, the SFLS model as an approximation and the proposed discretization scheme. Section \ref{section: tests} provides the numerical results and comparisons between the proposed method and existing models. Finally, we conclude the paper in Section \ref{section: conclusion}.

\section{Methodology}
\label{section: methodology}

\subsection{Heat extraction rate and wall temperature of borehole}

Consider a three-dimensional infinite isotropic medium with uniform thermal properties including thermal conductivity $k$ and thermal diffusivity $\alpha$. An instantaneous point heat source extracts energy $Q$ at the origin $(r=0)$, then the temperature change induced by this heat source at arbitrary point is dependent on the elapsed time $t$  and the distance to the heat source $d$, given by the product of $Q$ and Green function $K(d,t)$ \cite{CarslawJaeger1959}.

\begin{equation}
    \Delta T(d,t)=-QK( d,t )
\end{equation}

\begin{equation}
    K(d,t) = \frac{1}{\rho c (4\pi \alpha t)^{\frac{3}{2}}}\exp \left(-\frac{d^2}{4 \alpha t}\right)
\end{equation}

The integral of $K(d,t)$ over time from $t_1$ to $t_2$ is provided for reference where $\text{erfc}$  is the complementary error function.

\begin{equation}
    J(d,t_1,t_2) = \int_{t_1}^{t_2} K(d,\tau) \ d\tau = \frac{1}{4\pi kd}\left(\text{erfc}\left(\frac{d}{2\sqrt{\alpha t_2}}\right) - \text{erfc}\left(\frac{d}{2\sqrt{\alpha t_1}}\right)\right)
\end{equation}

\begin{equation}
    J(d,0,t_2) = \int_{t_1}^{t_2} K(d,\tau) \ d\tau = \frac{1}{4\pi kd}\text{erfc}\left(\frac{d}{2\sqrt{\alpha t_2}}\right)
\end{equation}

\begin{equation}
    \text{erfc}(x)=\frac{2}{\sqrt{\pi}}\int_{x}^{\infty}e^{-t^2}dt
\end{equation}

\begin{figure*}[!htb]
    \centering
    \includegraphics{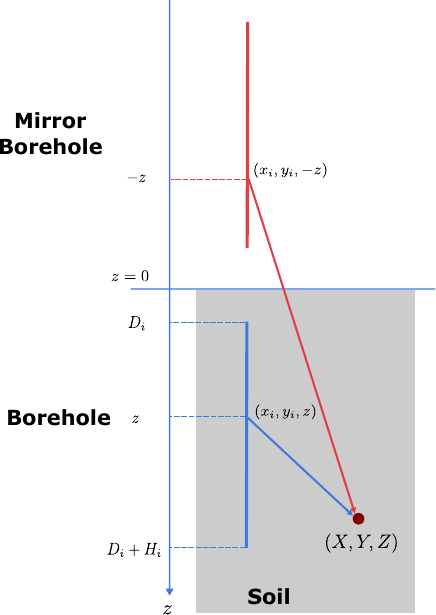}
    \caption{Spatial superposition of boreholes and their mirrors}
    \label{fig: borehole}
\end{figure*}

For a bore field composed of multiple vertical boreholes, we assume there are $N_b$ vertical boreholes located at positions $(x_i, y_i)$ for $i = 1, 2, \ldots, N_b$ and  with buried depth $D_i$ and length $H_i$. The total temperature change at a point $(x, y, z)$ due to all boreholes can be obtained by summing the contributions from each borehole. Each borehole is regarded as a line composed of many point heat sources, and the heat extraction and wall temperature of each borehole varies along the depth in ground and time, so we assume that the heat extraction rate per length and wall temperature is a bivariate function with respect to both depth and time expressed as $Q(z,t)$ and $T(z,t)$, respectively. As shown in Figure \ref{fig: borehole}, we introduce mirror boreholes with opposite heat extraction rate above the ground to maintain constant temperature at the ground surface as $T_g$ , namely the undisturbed ground temperature.

The temperature at arbitrary point $(X,Y,Z)$ and a specific time $t$ is given by the temporal and spatial superposition:

\begin{equation}
    \begin{split}
        T(X,Y,Z,t) = T_g - \sum_{i=1}^{N_b} \int_{D_i}^{D_i+H_i} \int_{0}^{t} Q_i(z,\tau)K(r(X-x_i, Y-y_i, Z-z),t-\tau) \ d\tau dz \\
        + \sum_{i=1}^{N_b} \int_{D_i}^{D_i+H_i} \int_{0}^{t} Q_i(z,\tau)K(r(X-x_i, Y-y_i, Z+z),t-\tau) \ d\tau dz
    \end{split}
\end{equation}

\begin{equation}
    r(x,y,z) = \sqrt{\max(x^2+y^2,r_b^2) + z^2}
\end{equation}

If the heat extraction rate is discretized into $T_{max}$ slots at points $t_0=0,t_1,t_2,\ldots,t_{T_{max}}$, we notice that:

\begin{equation}
    \begin{split}
        \int_{0}^{t_k} Q_i(z,\tau)K(d,t-\tau) \ d\tau & = \sum_{p=1}^{k} Q_i(z,t_p) \int_{t_{p-1}}^{t_p} K(d,t-\tau) \ d\tau   \\
                                                      & = \sum_{p=1}^{k} Q_i(z,t_p) \int_{t-t_p}^{t-t_{p-1}} K(d,\tau) \ d\tau \\
                                                      & = \sum_{p=1}^{k} Q_i(z,t_p) J(d,t-t_p,t-t_{p-1})
    \end{split}
\end{equation}

So, we have:

\begin{equation}
    \begin{split}
        T(X,Y,Z,t_k) = T_g - \sum_{p=1}^{k} \sum_{i=1}^{N_b} \int_{D_i}^{D_i+H_i} Q_i(z,t_p)J(r(X-x_i, Y-y_i, Z-z),t_k-t_p,t_k-t_{p-1}) \ dz \\
        + \sum_{p=1}^{k} \sum_{i=1}^{N_b} \int_{D_i}^{D_i+H_i} Q_i(z,t_p)J(r(X-x_i, Y-y_i, Z+z),t_k-t_p,t_k-t_{p-1}) \ dz
    \end{split}
\end{equation}

The $\displaystyle\sum_{p=1}^{k}$ summation means the temporal superposition and $\displaystyle\sum_{i=1}^{N_b} \displaystyle\int_{D_i}^{D_i+H_i}$ denotes the spatial superposition from all boreholes and their mirrors. The contribution of each borehole is written as an integral along its depth to accumulate the influence of varying heat extraction rate. $r$ is the distance from target point $(X,Y,Z)$  to the source point $(x_i,y_i,z)$ and its mirror $(x_i,y_i,-z)$. $r_b$ is the radius of each borehole to avoid singularity when the target point is on the wall of borehole ($x=x_i$  and  $y=y_i$). If the coordinates $(X,Y,Z)$ is set at the borehole wall $(x_i,y_i,Z)$, this formula gives the borehole wall temperature.

We separate the expression of temperature into two parts to discriminate the contributions from previous time-steps and current time-step:

\begin{equation}
    \label{eq: T1}
    \begin{split}
        \MoveEqLeft T(X,Y,Z,t_k) = T_0(X,Y,Z,t_k)                                                       \\
         & - \sum_{i=1}^{N_b} \int_{D_i}^{D_i+H_i} Q_i(z,t_k)J(r(X-x_i, Y-y_i, Z-z),0,t_k-t_{k-1}) \ dz \\
         & + \sum_{i=1}^{N_b} \int_{D_i}^{D_i+H_i} Q_i(z,t_k)J(r(X-x_i, Y-y_i, Z+z),0,t_k-t_{k-1}) \ dz
    \end{split}
\end{equation}

\begin{equation}
    \label{eq: T2}
    \begin{split}
        \MoveEqLeft T_0(X,Y,Z,t_k) = T_g                                                                                       \\
         & - \sum_{p=1}^{k-1} \sum_{i=1}^{N_b} \int_{D_i}^{D_i+H_i} Q_i(z,t_p)J(r(X-x_i, Y-y_i, Z-z),t_k-t_p,t_k-t_{p-1}) \ dz \\
         & + \sum_{p=1}^{k-1} \sum_{i=1}^{N_b} \int_{D_i}^{D_i+H_i} Q_i(z,t_p)J(r(X-x_i, Y-y_i, Z+z),t_k-t_p,t_k-t_{p-1}) \ dz
    \end{split}
\end{equation}

In the context of $g$-function (thermal response function), the average heat extraction rate per length is constant:

\begin{equation}
    \label{eq: Constant heat}
    Q(t_k) = \frac{\displaystyle\sum_{i=1}^{N_b} \int_{D_i}^{D_i+H_i} Q_i(z,t_k) \ dz}{\displaystyle\sum_{i=1}^{N_b} H_i} = 1
\end{equation}

As discussed in \cite{cimmino2014semi}, we need extra boundary conditions to solve the value of $g$-function, and the uniform borehole wall temperature (UBWT) boundary condition is chosen in this paper, which assumes that all the borehole walls have the same time-varying temperatures:

\begin{equation}
    \label{eq: UBWT}
    T(x_i,y_i,z,t_k) = T(t_k) (i=1,2,\ldots,N_b, D_i \leq z \leq D_i+H_i)
\end{equation}

\ref{eq: T1}, \ref{eq: T2}, \ref{eq: Constant heat} and \ref{eq: UBWT} are the governing integral equations of bore field that describes the spatio-temporal evolution of heat extraction rate and borehole wall temperature under UBWT boundary condition. Our goal is to obtain the spatial distribution and temporal evolution of the heat exchange rate $Q_i(z,t_k)$ and uniform temperature of the borehole wall $T(t_k)$.

If we recall the definition of $g$-function, it is obvious that $g$-function at time $t_k$ is directly related to the uniform borehole wall temperature if equation \ref{eq: Constant heat} holds.

\begin{equation}
    T(t) = T_g - \frac{1}{2 \pi k} \int_{0}^{t} Q(t-\tau) \frac{dg}{dt}(\tau) \ d\tau = T_g - \frac{1}{2 \pi k} g(t)
\end{equation}

In the next section, we will discuss how to solve these integral equations to acquire the $g$-function of bore field under UBWT boundary condition.

\textbf{Note}: The equations can be solved from $t_1$ to $t_{T_{max}}$ recursively. If we know the distribution of heat extraction rate per length of each bore hole $Q_i(z,t)$ at all time-points in front of $t_k$ (including $t_k$), we can compute $T_0(X,Y,Z,t_{k+1})$ according to \ref{eq: T2}. Thus, the key is to solve equation \ref{eq: T1} for each time-step.

\subsection{Stacked finite line source model as approximate solution to integral equations}

We notice that integral equations \ref{eq: T1}, \ref{eq: T2} and \ref{eq: Constant heat} all contain the spatial integral of the heat extraction rate $Q_i(z,t_k)$ as an unknown function with respect to the depth $z$. It is difficult to obtain the analytical form of $Q_i(z,t_k)$ as a continuous function, so we need to find an good approximation.

\begin{figure*}[!htb]
    \centering
    \includegraphics{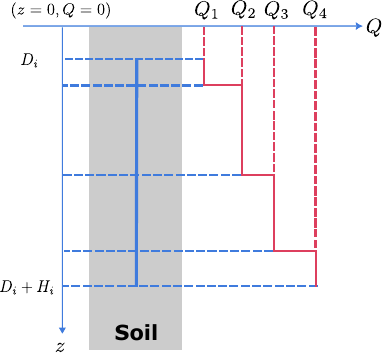}
    \caption{Stacked finite line source model}
    \label{fig: sfls}
\end{figure*}

Stacked finite line source (SFLS) model divides each borehole into segments and approximates $Q_i(z,t_k)$ as a stepwise function that remains constant in each segment. If each borehole is divided into $N_s$ segments, the integral of heat extraction can be decomposed into summation of $N_s$ discrete terms:
\begin{equation}
    \begin{split}
        \MoveEqLeft T(X,Y,Z,t_k) = T_0(X,Y,Z,t_k)                                                                                 \\
         & - \sum_{i=1}^{N_b} \sum_{j=1}^{N_s} Q_{i,j}(t_k) \int_{LB_{i,j}}^{UB_{i,j}} J(r(X-x_i, Y-y_i, Z-z),0,t_k-t_{k-1}) \ dz \\
         & + \sum_{i=1}^{N_b} \sum_{j=1}^{N_s} Q_{i,j}(t_k) \int_{LB_{i,j}}^{UB_{i,j}} J(r(X-x_i, Y-y_i, Z+z),0,t_k-t_{k-1}) \ dz
    \end{split}
\end{equation}
\begin{equation}
    \begin{split}
        \MoveEqLeft T_0(X,Y,Z,t_k) = T_g                                                                                                                \\
         & - \sum_{p=1}^{k-1} \sum_{i=1}^{N_b} \sum_{j=1}^{N_s} Q_{i,j}(t_p) \int_{LB_{i,j}}^{UB_{i,j}}J(r(X-x_i, Y-y_i, Z-z),t_k-t_p,t_k-t_{p-1}) \ dz \\
         & + \sum_{p=1}^{k-1} \sum_{i=1}^{N_b} \sum_{j=1}^{N_s} Q_{i,j}(t_p) \int_{LB_{i,j}}^{UB_{i,j}}J(r(X-x_i, Y-y_i, Z+z),t_k-t_p,t_k-t_{p-1}) \ dz
    \end{split}
\end{equation}
where $LB_{i,j}$ and $UB_{i,j}$ are the lower and upper bounds of the $j$-th segment on the $i$-th borehole, and $Q_{i,j}(t_k)$ is the average heat extraction rate per length of the $j$-th segment on the $i$-th borehole at time $t_k$.

The UBWT boundary condition \ref{eq: UBWT} is approximated by constraining the average temperature of each segment to be equal. For the $n$-th segment on the $m$-th borehole, its average temperature is the same with the uniform wall temperature $T(t_k)$:
\begin{equation}
    \label{eq: segment UBWT}
    T(t_k) = \frac{\displaystyle\int_{LB_{m,n}}^{UB_{m,n}} T(x_m,y_m,z,t_k) \ dz}{UB_{m,n}-LB_{m,n}}
\end{equation}

\begin{equation}
    \label{eq: UBWT_segment}
    \begin{split}
        \MoveEqLeft (UB_{m,n}-LB_{m,n})T(t_k) = (UB_{m,n}-LB_{m,n})T_g                                                                                                                              \\
         & - \sum_{p=1}^{k-1} \sum_{i=1}^{N_b} \sum_{j=1}^{N_s} Q_{i,j}(t_p) \int_{LB_{m,n}}^{UB_{m,n}} \int_{LB_{i,j}}^{UB_{i,j}}J(r(x_m-x_i, y_m-y_i, z_2-z_1),t_k-t_p,t_k-t_{p-1}) \ dz_1 \ dz_2 \\
         & + \sum_{p=1}^{k-1} \sum_{i=1}^{N_b} \sum_{j=1}^{N_s} Q_{i,j}(t_p) \int_{LB_{m,n}}^{UB_{m,n}} \int_{LB_{i,j}}^{UB_{i,j}}J(r(x_m-x_i, y_m-y_i, z_2+z_1),t_k-t_p,t_k-t_{p-1}) \ dz_1 \ dz_2 \\
         & - \sum_{i=1}^{N_b} \sum_{j=1}^{N_s} Q_{i,j}(t_k) \int_{LB_{m,n}}^{UB_{m,n}} \int_{LB_{i,j}}^{UB_{i,j}} J(r(x_m-x_i, y_m-y_i, z_2-z_1),0,t_k-t_{k-1}) \ dz_1 \ dz_2                       \\
         & + \sum_{i=1}^{N_b} \sum_{j=1}^{N_s} Q_{i,j}(t_k) \int_{LB_{m,n}}^{UB_{m,n}} \int_{LB_{i,j}}^{UB_{i,j}} J(r(x_m-x_i, y_m-y_i, z_2+z_1),0,t_k-t_{k-1}) \ dz_1 \ dz_2
    \end{split}
\end{equation}

The double integral appearing in \ref{eq: UBWT_segment} is called segment-to-segment response factor that reflects how the heat extraction of the $j$-th segment on the $i-$th borehole influences the wall temperature of the $n$-th segment on the $m$-th borehole. These factors are denoted as:

\begin{equation}
    \begin{split}
        \MoveEqLeft R^+_{(i,j)\to(m,n)}(t_1,t_2) = \int_{LB_{m,n}}^{UB_{m,n}} \int_{LB_{i,j}}^{UB_{i,j}}J(r(x_m-x_i, y_m-y_i, z_2-z_1),t_1, t_2) \ dz_1 \ dz_2
    \end{split}
\end{equation}

\begin{equation}
    \begin{split}
        \MoveEqLeft R^-_{(i,j)\to(m,n)}(t_1,t_2) = \int_{LB_{m,n}}^{UB_{m,n}} \int_{LB_{i,j}}^{UB_{i,j}} J(r(x_m-x_i, y_m-y_i, z_2+z_1),t_1,t_2) \ dz_1 \ dz_2
    \end{split}
\end{equation}

According to \cite{claesson2011analytical}, the segment-to-segment response factor can be expressed as a single integral as follows:

\begin{equation}
    \label{eq: segment-to-segment response factor}
    \begin{split}
        \MoveEqLeft \int_{A}^{B} \int_{C}^{D} J(\sqrt{r^2+(z_2-z_1)^2},t_1,t_2) \ dz_1 \ dz_2                                                                                          \\
         & = \int_{t_1}^{t_2} \int_{A}^{B} \int_{C}^{D} \frac{1}{\rho c (4\pi \alpha \tau)^{\frac{3}{2}}}\exp\left(-\frac{r^2+(z_2-z_1)^2}{4 \alpha \tau}\right) \ dz_1 \ dz_2 \ d\tau \\
         & = \frac{1}{4\pi k} \int_{1/\sqrt{4 \alpha t_2}}^{1/\sqrt{4 \alpha t_1}} \frac{1}{s^2} \exp(-r^2s^2) [ierf((B-C)s)                                                           \\
         & - ierf((A-C)s) + ierf((A-D)s) - ierf((B-D)s)]\ ds
    \end{split}
\end{equation}

\begin{equation}
    \text{erf}(x)=\frac{2}{\sqrt{\pi}}\int_{0}^{x}e^{-t^2} \ dt
\end{equation}

\begin{equation}
    ierf(x)=\int_{0}^{x}\text{erf}(t) \ dt=x\text{erf}(x)-\frac{1}{\sqrt{\pi}}(1-e^{-x^2})
\end{equation}

Then we can rewrite equation \ref{eq: UBWT_segment} as:

\begin{equation}
    \label{eq: SFLS temp}
    \begin{split}
        \MoveEqLeft (UB_{m,n}-LB_{m,n})T(t_k) = (UB_{m,n}-LB_{m,n})T_g                                                                                             \\
         & - \sum_{p=1}^{k-1} \sum_{i=1}^{N_b} \sum_{j=1}^{N_s} Q_{i,j}(t_p) (R^+_{(i,j)\to(m,n)}(t_k-t_p,t_k-t_{p-1}) - R^-_{(i,j)\to(m,n)}(t_k-t_p,t_k-t_{p-1})) \\
         & - \sum_{i=1}^{N_b} \sum_{j=1}^{N_s} Q_{i,j}(t_k) (R^+_{(i,j)\to(m,n)}(0,t_k-t_{k-1}) - R^-_{(i,j)\to(m,n)}(0,t_k-t_{k-1}))
    \end{split}
\end{equation}

The average heat extraction rate equation is written as:
\begin{equation}
    \label{eq: SFLS heat}
    \begin{split}
        \MoveEqLeft \frac{\displaystyle\sum_{i=1}^{N_b} \sum_{j=1}^{N_s} (UB_{i,j}-LB_{i,j}) Q_{i,j}(t_k)}{\displaystyle\sum_{i=1}^{N_b} H_i} = 1
    \end{split}
\end{equation}

Combining $N_bN_s$ equations \ref{eq: SFLS temp} and the single equation \ref{eq: SFLS heat} as a linear system, we can solve $N_bN_s+1$  unknown variables: the average heat extraction rate of each segment $Q_{i,j}(t_k)(1 \le i \le N_b, 1\le j \le N_s)$ and the uniform borehole wall temperature $T(t_k)$:

\begin{equation}
    \label{eq: Linear system}
    \begin{bmatrix}
        \mathbf{A}(0,t_k-t_{k-1}) & \mathbf{1} \\
        \mathbf{c}^T              & 0
    \end{bmatrix}
    \begin{bmatrix}
        \mathbf{Q}(t_k) \\
        T(t_k)
    \end{bmatrix}
    =
    \begin{bmatrix}
        \mathbf{1}T_g - \displaystyle\sum_{p=1}^{k-1} \mathbf{A}(t_k-t_p,t_k-t_{p-1}) \mathbf{Q}(t_p) \\
        1
    \end{bmatrix}
\end{equation}

Therefore, we have derived the core equations of SFLS model under UBWT boundary condition from integral equations that have the same form with \cite{cimmino2014semi}. As a discrete approximation of the integral equations, the SFLS model has the following inherent limitations:

\begin{enumerate}
    \item The stepwise approximation of heat extraction rate is discontinuous and not accurate enough to fit the actual continuous spatial distribution of heat flux on borehole wall.
    \item The numerical simulation has proved that the heat extraction rate experiences rapid change near the heads and tails of borehole and remains flat at the middle of borehole \cite{cimmino2014semi,cook2021development}, so nonuniform segmentation is required to arrange more segments on both terminals of boreholes to capture the rapid change and reduce the error of approximation. However, there is no theoretical guidance on how to select the appropriate discretization scheme.
    \item Calculating the segment-to-segment response factors for all pairs of segments as an integral is time consuming and occupies most of the time in computation of $g$-functions.
\end{enumerate}

\subsection{Novel discretization method based on numeric quadrature rule}

In this paper, we adopt the Nyström method \cite{hackbusch2012integral} to solve the integral equation, which is a numerical method to approximate the integral operator with a quadrature rule. For example, the following one-dimensional integral can be approximated by a weighted sum of function values at some specific points:
\begin{equation}
    \int_{a}^{b} f(x) \ dx \approx \sum_{i=1}^{n} w_i f(x_i)
\end{equation}
where $x_i$ are the quadrature points and $w_i$ are the corresponding weights. The accuracy of the approximation depends on the number of quadrature points and the selection of quadrature rule.

For the integral equations of boreholes, the spatial integral in \ref{eq: T1}, \ref{eq: T2} and \ref{eq: Constant heat} are replaced with a weighted sum with respect to the heat extraction rate per length at some specific depths.

\begin{equation}
    \label{eq: T1 quad}
    \begin{split}
        \MoveEqLeft T(X,Y,Z,t_k) = T_0(X,Y,Z,t_k)                                                                                                \\
         & - \sum_{i=1}^{N_b} \sum_{j=1}^{N_{quad}^{heat}} w_{ij}^{heat} Q_i(z_{ij}^{heat},t_k)J(r(X-x_i, Y-y_i, Z-z_{ij}^{heat}),0,t_k-t_{k-1}) \\
         & + \sum_{i=1}^{N_b} \sum_{j=1}^{N_{quad}^{heat}} w_{ij}^{heat} Q_i(z_{ij}^{heat},t_k)J(r(X-x_i, Y-y_i, Z+z_{ij}^{heat}),0,t_k-t_{k-1})
    \end{split}
\end{equation}

\begin{equation}
    \label{eq: T2 quad}
    \begin{split}
        \MoveEqLeft T_0(X,Y,Z,t_k) = T_g                                                                                                                                 \\
         & - \sum_{p=1}^{k-1} \sum_{i=1}^{N_b} \sum_{j=1}^{N_{quad}^{heat}} w_{ij}^{heat} Q_i(z_{ij}^{heat},t_p) J(r(X-x_i, Y-y_i, Z-z_{ij}^{heat}),t_k-t_p,t_k-t_{p-1}) \\
         & + \sum_{p=1}^{k-1} \sum_{i=1}^{N_b}\sum_{j=1}^{N_{quad}^{heat}} w_{ij}^{heat} Q_i(z_{ij}^{heat},t_p) J(r(X-x_i, Y-y_i, Z+z_{ij}^{heat}),t_k-t_p,t_k-t_{p-1})
    \end{split}
\end{equation}

\begin{equation}
    \label{eq: Constant heat quad}
    \begin{split}
        \frac{\displaystyle\sum_{i=1}^{N_b} \sum_{j=1}^{N_{quad}^{heat}} w_{ij}^{heat} Q_i(z_{ij}^{heat},t_k)}{\displaystyle\sum_{i=1}^{N_b} H_i} = 1
    \end{split}
\end{equation}

The weight $w_{ij}^{heat}$ and points $z_{ij}^{heat}$ can be determined by arbitrary quadrature rule and we choose Gauss-Legendre rule in this paper. The number of quadrature points $N_{quad}^{heat}$ is a parameter that can be adjusted to control the accuracy of the approximation.

\begin{figure*}[!htb]
    \centering
    \includegraphics[scale=0.4]{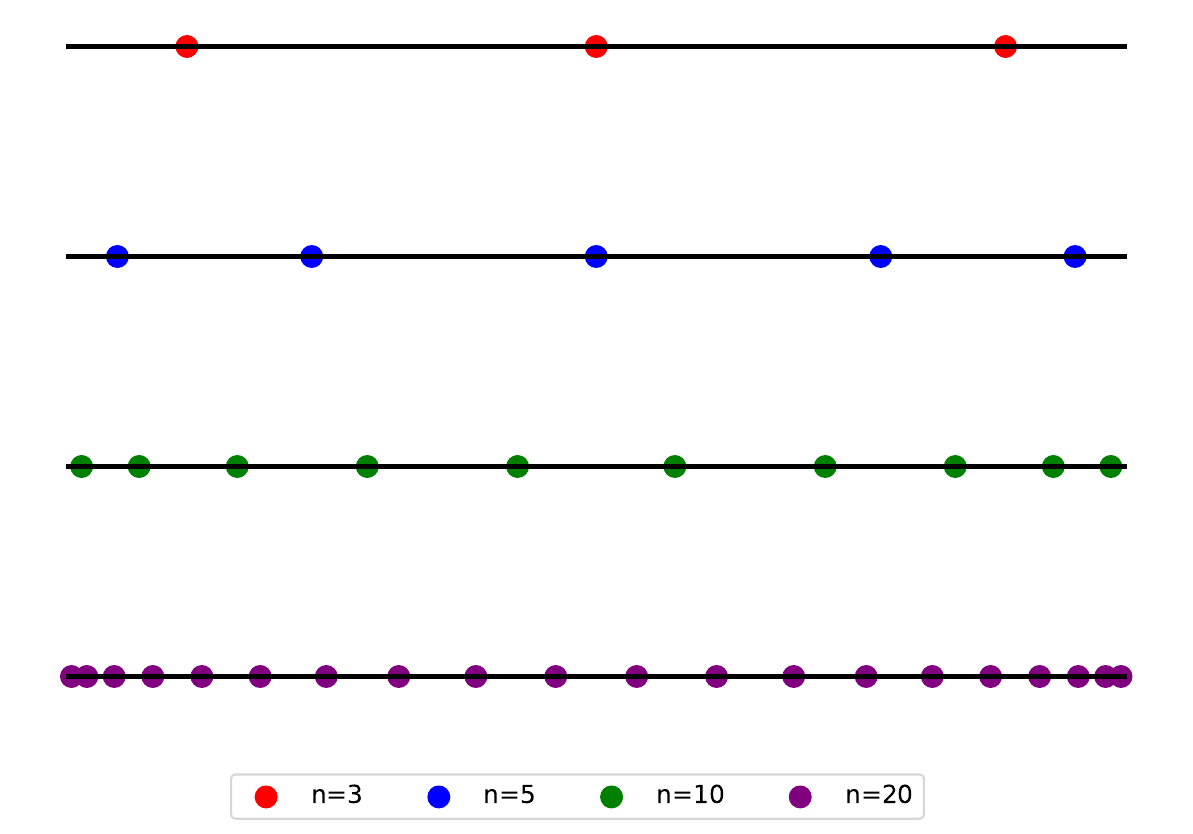}
    \caption{Quadrature points of Gauss-Legendre rule with different order in an interval}
    \label{fig: gaussquad}
\end{figure*}

As shown in Figure \ref{fig: gaussquad}, the quadrature points are distributed in the interval nonuniformly with corresponding weights. As the order of quadrature rule increases, the quadrature points are more dense to reduce the error of integral. The locations and weights of quadrature points  are pre-calculated and stored in advance.

Similar with the SFLS model, the UBWT boundary condition \ref{eq: UBWT} is approximated by constraining the average temperature of each segment of borehole to be equal. In order to solve $N_bN_{quad}^{heat}$ variables $Q(z_{ij}^{heat},t_k)(1 \le i \le N_b, 1\le j \le N_{quad}^{heat})$, we seperate each borehole into $N_{quad}^{heat}$ segments by the heads and tails of borehole and the midpoints of quadrature points. This segmentation scheme ensures that there is one quadrature point in each segment.

\begin{equation}
    LB_{i,j} =
    \begin{cases}
        D_i                                          & j=1                     \\
        \displaystyle\frac{1}{2}(z_{(i-1)j}+z_{ij}), & 1<j \le N_{quad}^{heat}
    \end{cases}
\end{equation}

\begin{equation}
    UB_{i,j} =
    \begin{cases}
        \displaystyle\frac{1}{2}(z_{ij}+z_{(i+1)j}), & 1\le j < N_{quad}^{heat} \\
        D_i + H_i                                    & j=N_{quad}^{heat}
    \end{cases}
\end{equation}

After the segmentation is determined, the integral in \ref{eq: segment UBWT} is also approximated by a weighted sum of temperatures at quadrature points in each segment according to the same Gauss-Legendre quadrature rule:
\begin{equation}
    \label{eq: segment UBWT quad}
    \begin{split}
        T(t_k) & = \frac{\displaystyle\int_{LB_{m,n}}^{UB_{m,n}} T(x_m,y_m,z,t_k) \ dz}{UB_{m,n}-LB_{m,n}}               \\
               & = \frac{1}{UB_{m,n}-LB_{m,n}} \sum_{l=1}^{N_{quad}^{temp}} w_{mnl}^{temp} T(x_m,y_m,z_{mnl}^{temp},t_k)
    \end{split}
\end{equation}
where $z_{mnl}^{temp}$ are the quadrature points in the $n$-th segment on the $m$-th borehole and $w_{mnl}^{temp}$ are the corresponding weights. $N_{quad}^{temp}$ is the number of quadrature points to approximate the integral of temperature in each segment.

Substituting the expression of temperature \ref{eq: T1 quad} and \ref{eq: T2 quad} into \ref{eq: segment UBWT quad}, we have:

\begin{equation}
    \label{eq: quad temp}
    \begin{split}
        \MoveEqLeft (UB_{m,n}-LB_{m,n})T(t_k) = (UB_{m,n}-LB_{m,n})T_g                                                                                                                                                                \\
         & - \sum_{p=1}^{k-1} \sum_{i=1}^{N_b} \sum_{j=1}^{N_{quad}^{heat}} w_{ij}^{heat} Q_i(z_{ij}^{heat},t_p) \sum_{l=1}^{N_{quad}^{temp}} w_{mnl}^{temp} J(r(x_m-x_i, y_m-y_i, z_{mnl}^{temp}-z_{ij}^{heat}),t_k-t_p,t_k-t_{p-1}) \\
         & + \sum_{p=1}^{k-1} \sum_{i=1}^{N_b} \sum_{j=1}^{N_{quad}^{heat}} w_{ij}^{heat} Q_i(z_{ij}^{heat},t_p) \sum_{l=1}^{N_{quad}^{temp}} w_{mnl}^{temp} J(r(x_m-x_i, y_m-y_i, z_{mnl}^{temp}+z_{ij}^{heat}),t_k-t_p,t_k-t_{p-1}) \\
         & - \sum_{i=1}^{N_b} \sum_{j=1}^{N_{quad}^{heat}} w_{ij}^{heat} Q_i(z_{ij}^{heat},t_k) \sum_{l=1}^{N_{quad}^{temp}} w_{mnl}^{temp} J(r(x_m-x_i, y_m-y_i, z_{mnl}^{temp}-z_{ij}^{heat}),0,t_k-t_{k-1})                        \\
         & + \sum_{i=1}^{N_b} \sum_{j=1}^{N_{quad}^{heat}} w_{ij}^{heat} Q_i(z_{ij}^{heat},t_k) \sum_{l=1}^{N_{quad}^{temp}} w_{mnl}^{temp} J(r(x_m-x_i, y_m-y_i, z_{mnl}^{temp}+z_{ij}^{heat}),0,t_k-t_{k-1})
    \end{split}
\end{equation}

Combining $N_bN_{quad}^{heat}$ equations \ref{eq: quad temp} and the single equation \ref{eq: Constant heat quad}, we can obtain a linear system with the similar structure with \ref{eq: Linear system} to solve $N_bN_{quad}^{heat}+1$  unknown variables: heat extraction rate of each quadrature point $Q_i(z_{ij}^{heat},t_k)(1 \le i \le N_b, 1\le j \le N_{quad}^{heat})$ and the uniform borehole wall temperature $T(t_k)$:

Compared with the SFLS model, the proposed formulation has the following advantages:

\begin{enumerate}
    \item We do not assume the stepwise structure of heat extraction rate $Q_i(z,t_k)$ and only uses limited quadrature points to approximate the integral with numerical quadrature rule, which ensures higher accuracy with the same number of discretization points.
    \item The discretization scheme to approximate the heat extraction rate and borehole wall temperature are automatically generated by Gauss-Legendre quadrature rule after the number of quadrature points $N_{quad}^{heat}$ and $N_{quad}^{temp}$ are determined, so it is universal for all configurations of bore field and has theoretical guarantee on the accuracy of approximation.
    \item We only need to compute a series of point-to-point response factors in the form of $J(d,t_1,t_2)$ to formulate the discretized equations. It is analytical and much easier to compute than the segment-to-segment response factor as integral in \ref{eq: segment-to-segment response factor}.
\end{enumerate}

\subsection{Ill-conditioned discretized equations and regularization}
\label{sec: regularization}

The original integral equations including \ref{eq: T1}, \ref{eq: T2}, \ref{eq: Constant heat} and \ref{eq: UBWT} can be reduced to the Fredholm integral equation of the first kind:
\begin{equation}
    \label{eq: Fredholm}
    \int_{a}^{b} K(x,y)f(y) \ dy = g(x)
\end{equation}
where $K(x,y)$ is the kernel function, $f(y)$ is the unknown function to be solved, and $g(x)$ is the known boundary condition. In our case, the kernel function $K(x,y)$ is the response factor $J(d,t_1,t_2)$, the unknown function $f(y)$ is the heat extraction rate $Q_i(z,t_k)$ and the known function $g(x)$ is the uniform borehole wall temperature $T(t_k)$.

The Fredholm integral equation of the first kind is a well-known ill-posed problem \cite{wazwaz2011linear} where the solution is not unique and sensitive to the perturbation of data. Thus, the discretization of the integral equation in form of linear system \ref{eq: Linear system} is ill-conditioned, and we have observed that its condition number increases quickly with the order of discretization. The ill-conditioned discretized equations lead to numerical instability and incorrect values of $g$-function, which is confirmed and discussed in the numerical tests.

We use the regularization method proposed in \cite{wazwaz2011regularization} to transform the ill-posed Fredholm integral equation of the first kind into the well-defined second kind with an extra term:
\begin{equation}
    \label{eq: Fredholm kind 2}
    \int_{a}^{b} K(x,y)f(y) \ dy = g(x) - \lambda f(x)
\end{equation}
where $\lambda$ is a regularization parameter. When $\lambda$ is small, the solution of \ref{eq: Fredholm kind 2} is close to the solution of \ref{eq: Fredholm}.

The regularized form of \ref{eq: T1} is:
\begin{equation}
    T_{reg}(X,Y,Z,t_k) = T(X,Y,Z,t_k) - \lambda Q(X,Y,Z,t_k)
\end{equation}

Then the integral equation \ref{eq: segment UBWT} in SFLS model is turned into two parts:
\begin{equation}
    \begin{split}
        T(t_k) & = \frac{\displaystyle\int_{LB_{m,n}}^{UB_{m,n}} T_{reg}(x_m,y_m,z,t_k) \ dz}{UB_{m,n}-LB_{m,n}}                  \\
               & = \frac{\displaystyle\int_{LB_{m,n}}^{UB_{m,n}} T(x_m,y_m,z,t_k) \ dz}{UB_{m,n}-LB_{m,n}} - \lambda Q_{m,n}(t_k)
    \end{split}
\end{equation}

It is similar with the proposed discretization method:
\begin{equation}
    \begin{split}
        T(t_k) & = \frac{\displaystyle\int_{LB_{m,n}}^{UB_{m,n}} T_{reg}(x_m,y_m,z,t_k) \ dz}{UB_{m,n}-LB_{m,n}}                            \\
               & = \frac{\displaystyle\int_{LB_{m,n}}^{UB_{m,n}} T(x_m,y_m,z,t_k) \ dz}{UB_{m,n}-LB_{m,n}} - \lambda Q_m(z_{mn}^{heat},t_k)
    \end{split}
\end{equation}

The linear system \ref{eq: Linear system} is modified to the following form where the parameter $\lambda$ improves the ill-conditioned discretized equations by adding a small diagonal regularization to the coefficient matrix:

\begin{equation}
    \label{eq: Linear system after regularization}
    \begin{bmatrix}
        \mathbf{A}(0,t_k-t_{k-1}) + \lambda \mathbf{I} & \mathbf{1} \\
        \mathbf{c}^T                                   & 0
    \end{bmatrix}
    \begin{bmatrix}
        \mathbf{Q}(t_k) \\
        T(t_k)
    \end{bmatrix}
    =
    \begin{bmatrix}
        \mathbf{1}T_g - \displaystyle\sum_{p=1}^{k-1} \mathbf{A}(t_k-t_p,t_k-t_{p-1}) \mathbf{Q}(t_p) \\
        1
    \end{bmatrix}
\end{equation}

\section{Results}
\label{section: tests}

In this section, we will compare the proposed discretized formulation of $g$-function with the SFLS model in terms of the accuracy, convergence rate and computational efficiency.

Because the accuracy of SFLS model depends on the  segementation scheme to discretize the borehole, we choose three different segmentation schemes used in \cite{cimmino2024optimal} as the targets for comparison. The first one is the uniform segmentation scheme, which divides each borehole into $N_s$ segments with equal length. The second and third one are nonuniform segmentation schemes, which chooses the length of the smallest segments $h_1$ at both ends of the borehole and arranges the length of segments from the ends to middle of borehole as a geometric series with the common ratio $\gamma$. The second scheme chooses $\delta_1=\frac{h_1}{H}=0.02\times\frac{8}{N_s}$ as a baseline and the third scheme chooses $\delta_1=\frac{h_1}{H}=0.005525\times\frac{8}{N_s}$ as the optimal nonuniform discretization trained on dataset including 557056 bore field configurations. The three segmentation schemes are called as ``uniform", ``nonuniform-base" and ``nonuniform-opt" in the rest of paper respectively.

The proposed formulation is implemented in pure Python with NumPy and SciPy libraries. The open source \texttt{pygfunction} module is used as the reference implementation of SFLS model with minimal modifications to add regularization and obtain the condition number of linear system. All numerical tests are performed on a laptop with Intel Core i7-1360P CPU and 32GB RAM.

\subsection{The convergence of $g$-function in a 2-borehole test case}

\begin{table}[]
    \centering
    \caption{Parameter of geothermal boreholes}
    \label{tab: parameter}
    \begin{tabular}{|l|l|}
        \hline
        Parameter                                     & Value              \\ \hline
        Buried depth $D$(m)                           & 4.0                \\
        Borehole length $H$(m)                        & 120.0              \\
        Radius of borehole $r_b$(m)                   & 0.1                \\
        Space between borehole(m)                     & 5.0                \\
        Thermal diffusivity of soil $\alpha$(m$^2$/s) & $1 \times 10^{-6}$ \\ \hline
    \end{tabular}
\end{table}

We choose a bore field of two boreholes as a small test case to evaluate the $g$-functions at ten time points from 2 years to 20 years with a time step of 2 years. The parameters are specified in Table \ref{tab: parameter}.

\subsubsection{Divergent $g$-functions without regularization}

In the first round of test, $g$-function is calculated by solving the original linear system \ref{eq: Linear system} without regularization.  The order of discretization is selected as follows:

\begin{itemize}
    \item The proposed discretization scheme with $N_{quad}^{heat}=[64, 128, 256, 512, 1024]$ and fixed $N_{quad}^{temp}=40$.
    \item The SFLS model with $N_s=[64, 128, 256, 512, 1024]$.
\end{itemize}

The results are shown in Tables \ref{tab: unreg-1} to \ref{tab: unreg-4} including the condition number of the linear system and the values of $g$-function at ten time points. We can observe that the value of $g$-function generally decreases and tends to converge to a stable value as the order of discretization increases from 64 to 1024. The exceptions happen when the order of discretization equals to 1024 for the proposed discretization scheme and the optimal nonuniform discretization scheme, where the value of $g$-function gradually deviates from the converged value and even becomes negative. This unrealistic result is caused by the condition number of the linear system increasing quickly with the order of discretization that leads to numerical instability and incorrect results. The biggest condition number reaches $5.19\times10^{18}$ and $2.27\times10^{15}$ and renders the solutions to these ill-conditioned linear systems unreliable because thay are very sensitive to the perturbation and floating point error accumulated in the right hand side of \ref{eq: Linear system} during computation. It contradicts the common intuition that more detailed discretization always provides more accurate value of $g$-function. Thus, the ill-conditioned problem we discussed in Section \ref{sec: regularization} is confirmed and the regularization method is necessary to improve the stability of solution.

\begin{table}[]
    \centering
    \caption{$g$-function of 2 boreholes given by proposed discretization without regularization}
    \label{tab: unreg-1}
    \begin{tabular}{|l|l|l|l|l|l|}
        \hline
        $N_{quad}^{heat}$ & 64     & 128    & 256    & 512                 & 1024                \\ \hline
        Condition number  & 294.8  & 352.6  & 69562  & $1.73\times10^{11}$ & $5.19\times10^{18}$ \\ \hline
        $g(t=2\text{a})$  & 5.5324 & 5.5309 & 5.5282 & 5.5257              & 5.5248              \\
        $g(t=4\text{a})$  & 6.1126 & 6.1107 & 6.1076 & 6.1049              & 6.1046              \\
        $g(t=6\text{a})$  & 6.4385 & 6.4364 & 6.4331 & 6.4302              & 6.4315              \\
        $g(t=8\text{a})$  & 6.6607 & 6.6584 & 6.6550 & 6.6519              & 6.6593              \\
        $g(t=10\text{a})$ & 6.8267 & 6.8243 & 6.8207 & 6.8176              & 6.8360              \\
        $g(t=12\text{a})$ & 6.9576 & 6.9551 & 6.9515 & 6.9482              & 6.9944              \\
        $g(t=14\text{a})$ & 7.0647 & 7.0621 & 7.0584 & 7.0551              & 7.1801              \\
        $g(t=16\text{a})$ & 7.1546 & 7.1519 & 7.1481 & 7.1448              & 7.4290              \\
        $g(t=18\text{a})$ & 7.2315 & 7.2287 & 7.2249 & 7.2215              & 7.8758              \\
        $g(t=20\text{a})$ & 7.2983 & 7.2955 & 7.2917 & 7.2882              & 9.2351              \\ \hline
    \end{tabular}
\end{table}

\begin{table}[]
    \centering
    \caption{$g$-function of 2 boreholes given by uniform discretization without regularization}
    \label{tab: unreg-2}
    \begin{tabular}{|l|l|l|l|l|l|}
        \hline
        $N_{s}$           & 64     & 128    & 256    & 512    & 1024   \\ \hline
        Condition number  & 13.29  & 17.27  & 41.00  & 151.4  & 1116   \\ \hline
        $g(t=2\text{a})$  & 5.5441 & 5.5409 & 5.5383 & 5.5361 & 5.5340 \\
        $g(t=4\text{a})$  & 6.1259 & 6.1221 & 6.1192 & 6.1166 & 6.1143 \\
        $g(t=6\text{a})$  & 6.4527 & 6.4486 & 6.4455 & 6.4428 & 6.4402 \\
        $g(t=8\text{a})$  & 6.6755 & 6.6712 & 6.6679 & 6.6651 & 6.6624 \\
        $g(t=10\text{a})$ & 6.8419 & 6.8375 & 6.8341 & 6.8311 & 6.8284 \\
        $g(t=12\text{a})$ & 6.9732 & 6.9686 & 6.9652 & 6.9622 & 6.9593 \\
        $g(t=14\text{a})$ & 7.0806 & 7.0759 & 7.0724 & 7.0693 & 7.0664 \\
        $g(t=16\text{a})$ & 7.1708 & 7.1660 & 7.1623 & 7.1592 & 7.1563 \\
        $g(t=18\text{a})$ & 7.2479 & 7.2430 & 7.2393 & 7.2362 & 7.2332 \\
        $g(t=20\text{a})$ & 7.3149 & 7.3100 & 7.3063 & 7.3031 & 7.3000 \\ \hline
    \end{tabular}
\end{table}

\begin{table}[]
    \centering
    \caption{$g$-function of 2 boreholes given by baseline nonuniform discretization without regularization}
    \label{tab: unreg-3}
    \begin{tabular}{|l|l|l|l|l|l|}
        \hline
        $N_{s}$           & 64     & 128    & 256    & 512    & 1024               \\ \hline
        Condition number  & 61.86  & 166.2  & 1811   & 175538 & $9.13\times10^{8}$ \\ \hline
        $g(t=2\text{a})$  & 5.5369 & 5.5348 & 5.5327 & 5.5306 & 5.5285             \\
        $g(t=4\text{a})$  & 6.1175 & 6.1151 & 6.1128 & 6.1104 & 6.1081             \\
        $g(t=6\text{a})$  & 6.4437 & 6.4411 & 6.4386 & 6.4361 & 6.4336             \\
        $g(t=8\text{a})$  & 6.6661 & 6.6634 & 6.6607 & 6.6581 & 6.6555             \\
        $g(t=10\text{a})$ & 6.8322 & 6.8294 & 6.8267 & 6.8240 & 6.8212             \\
        $g(t=12\text{a})$ & 6.9632 & 6.9603 & 6.9576 & 6.9548 & 6.9520             \\
        $g(t=14\text{a})$ & 7.0704 & 7.0675 & 7.0646 & 7.0618 & 7.0589             \\
        $g(t=16\text{a})$ & 7.1603 & 7.1574 & 7.1545 & 7.1516 & 7.1487             \\
        $g(t=18\text{a})$ & 7.2373 & 7.2343 & 7.2314 & 7.2284 & 7.2255             \\
        $g(t=20\text{a})$ & 7.3042 & 7.3011 & 7.2982 & 7.2952 & 7.2922             \\ \hline
    \end{tabular}
\end{table}

\begin{table}[]
    \centering
    \caption{$g$-function of 2 boreholes given by optimal nonuniform discretization without regularization}
    \label{tab: unreg-4}
    \begin{tabular}{|l|l|l|l|l|l|}
        \hline
        $N_{s}$           & 64     & 128    & 256                & 512                 & 1024                \\ \hline
        Condition number  & 765.6  & 16357  & $1.12\times10^{7}$ & $1.72\times10^{13}$ & $2.27\times10^{15}$ \\ \hline
        $g(t=2\text{a})$  & 5.5331 & 5.5310 & 5.5290             & 5.5267              & 5.5258              \\
        $g(t=4\text{a})$  & 6.1132 & 6.1109 & 6.1086             & 6.1060              & 6.1054              \\
        $g(t=6\text{a})$  & 6.4391 & 6.4366 & 6.4341             & 6.4314              & 6.4347              \\
        $g(t=8\text{a})$  & 6.6612 & 6.6586 & 6.6560             & 6.6532              & 6.6380              \\
        $g(t=10\text{a})$ & 6.8272 & 6.8245 & 6.8218             & 6.8189              & 6.8487              \\
        $g(t=12\text{a})$ & 6.9581 & 6.9553 & 6.9526             & 6.9496              & 6.8483              \\
        $g(t=14\text{a})$ & 7.0652 & 7.0623 & 7.0595             & 7.0565              & 7.4772              \\
        $g(t=16\text{a})$ & 7.1550 & 7.1521 & 7.1493             & 7.1462              & 5.8117              \\
        $g(t=18\text{a})$ & 7.2319 & 7.2290 & 7.2261             & 7.2230              & 10.8160             \\
        $g(t=20\text{a})$ & 7.2987 & 7.2958 & 7.2928             & 7.2897              & -3.5974             \\ \hline
    \end{tabular}
\end{table}

\subsubsection{Convergent $g$-functions with regularization}

\begin{figure*}[!htb]
    \centering
    \includegraphics[width=0.76\linewidth]{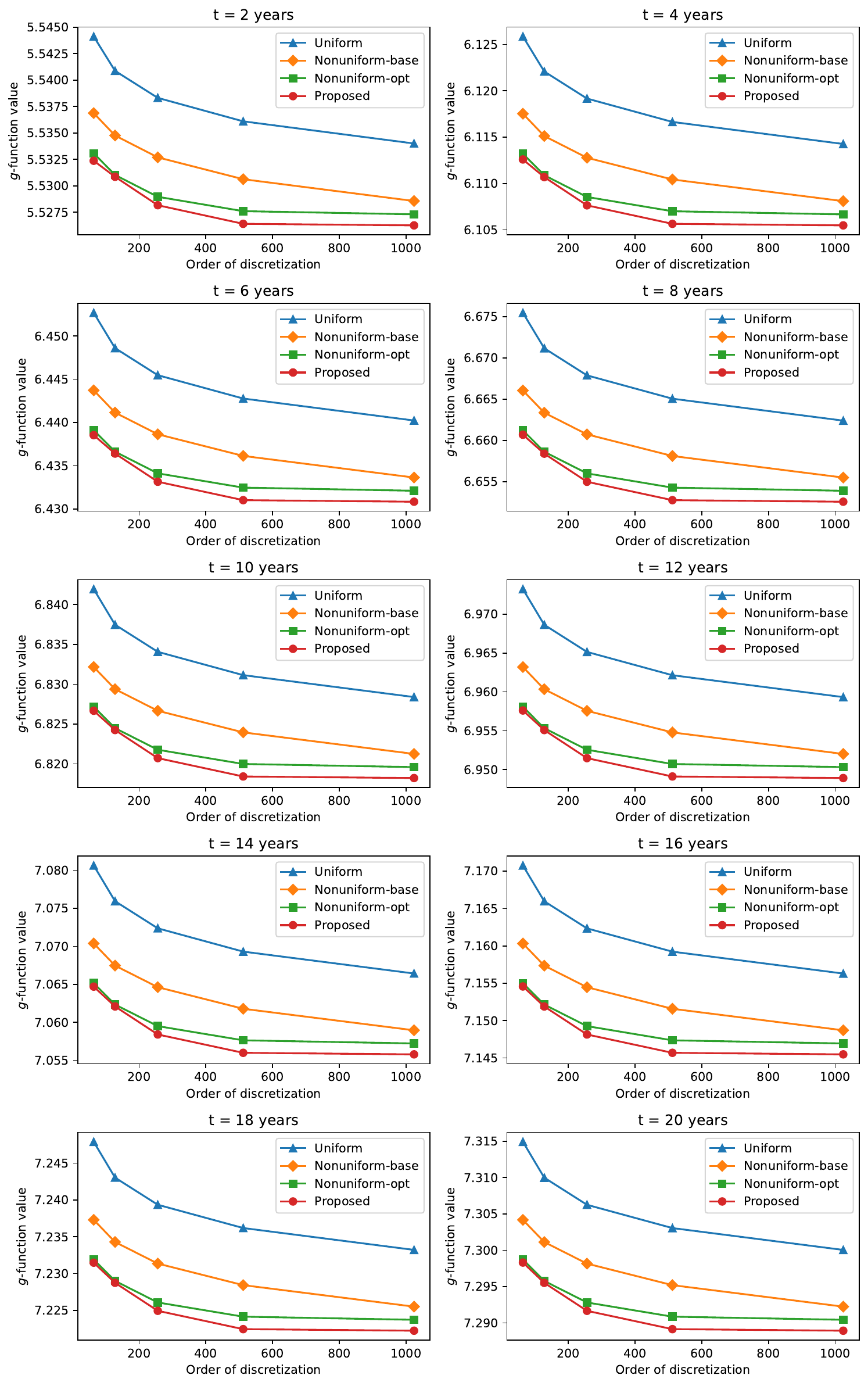}
    \caption{$g$-function of 2 boreholes given by different orders of discretization with regularization $\lambda=1\times 10^{-8}$}
    \label{fig: 2borehole}
\end{figure*}

In the second round of test, we apply the regularization method to the linear system \ref{eq: Linear system} with $\lambda=1\times 10^{-8}$ and solve the regularized linear system \ref{eq: Linear system after regularization}. The results are shown in Tables \ref{tab: reg-1} to \ref{tab: reg-4} and Figure \ref{fig: 2borehole}.

We observe that the highest condition numbers of the linear system when the order of discretization equals to 1024 are reduced from $5.19\times10^{18}$ to $1.48\times10^{10}$ after regularization, which indicates that the regularization method improves the stability of the solution effectively. As shown in Figure \ref{fig: 2borehole}, the values of $g$-function given by four methods all exhibit a clear trend to converge to a stable value as the order of discretization increases from 64 to 1024.

In terms of the convergence rate, it is obvious that the proposed discretization scheme converges faster than the other three methods including the optimal nonuniform discretization scheme in Figure \ref{fig: 2borehole}. The proposed method reaches stable values of $g$-function at $N_{quad}^{heat}=512$ and more accurate than both uniform and nonuniform discretizations.

\begin{table}[]
    \centering
    \caption{$g$-function of 2 boreholes given by proposed discretization with regularization}
    \label{tab: reg-1}
    \begin{tabular}{|l|l|l|l|l|l|}
        \hline
        $N_{quad}^{heat}$ & 64     & 128    & 256    & 512                & 1024                \\ \hline
        Condition number  & 294.8  & 352.6  & 69561  & $7.14\times10^{9}$ & $1.48\times10^{10}$ \\ \hline
        $g(t=2\text{a})$  & 5.5324 & 5.5309 & 5.5282 & 5.5264             & 5.5263              \\
        $g(t=4\text{a})$  & 6.1126 & 6.1107 & 6.1076 & 6.1057             & 6.1055              \\
        $g(t=6\text{a})$  & 6.4385 & 6.4364 & 6.4331 & 6.4310             & 6.4308              \\
        $g(t=8\text{a})$  & 6.6607 & 6.6584 & 6.6550 & 6.6528             & 6.6526              \\
        $g(t=10\text{a})$ & 6.8267 & 6.8243 & 6.8207 & 6.8184             & 6.8182              \\
        $g(t=12\text{a})$ & 6.9576 & 6.9551 & 6.9515 & 6.9491             & 6.9489              \\
        $g(t=14\text{a})$ & 7.0647 & 7.0621 & 7.0584 & 7.0560             & 7.0558              \\
        $g(t=16\text{a})$ & 7.1546 & 7.1519 & 7.1481 & 7.1457             & 7.1455              \\
        $g(t=18\text{a})$ & 7.2315 & 7.2287 & 7.2249 & 7.2225             & 7.2222              \\
        $g(t=20\text{a})$ & 7.2983 & 7.2955 & 7.2917 & 7.2892             & 7.2889              \\ \hline
    \end{tabular}
\end{table}

\begin{table}[]
    \centering
    \caption{$g$-function of 2 boreholes given by uniform discretization with regularization}
    \label{tab: reg-2}
    \begin{tabular}{|l|l|l|l|l|l|}
        \hline
        $N_{quad}^{heat}$ & 64     & 128    & 256    & 512    & 1024   \\ \hline
        Condition number  & 13.29  & 17.27  & 41.00  & 151.4  & 1116   \\ \hline
        $g(t=2\text{a})$  & 5.5441 & 5.5409 & 5.5383 & 5.5361 & 5.5340 \\
        $g(t=4\text{a})$  & 6.1259 & 6.1221 & 6.1192 & 6.1166 & 6.1143 \\
        $g(t=6\text{a})$  & 6.4527 & 6.4486 & 6.4455 & 6.4428 & 6.4402 \\
        $g(t=8\text{a})$  & 6.6755 & 6.6712 & 6.6679 & 6.6651 & 6.6624 \\
        $g(t=10\text{a})$ & 6.8419 & 6.8375 & 6.8341 & 6.8311 & 6.8284 \\
        $g(t=12\text{a})$ & 6.9732 & 6.9686 & 6.9652 & 6.9622 & 6.9593 \\
        $g(t=14\text{a})$ & 7.0806 & 7.0759 & 7.0724 & 7.0693 & 7.0664 \\
        $g(t=16\text{a})$ & 7.1708 & 7.1660 & 7.1623 & 7.1592 & 7.1563 \\
        $g(t=18\text{a})$ & 7.2479 & 7.2430 & 7.2393 & 7.2362 & 7.2332 \\
        $g(t=20\text{a})$ & 7.3149 & 7.3100 & 7.3063 & 7.3031 & 7.3000 \\ \hline
    \end{tabular}
\end{table}

\begin{table}[]
    \centering
    \caption{$g$-function of 2 boreholes given by baseline nonuniform discretization with regularization}
    \label{tab: reg-3}
    \begin{tabular}{|l|l|l|l|l|l|}
        \hline
        $N_{quad}^{heat}$ & 64     & 128    & 256    & 512    & 1024               \\ \hline
        Condition number  & 61.86  & 166.2  & 1811   & 175528 & $7.61\times10^{8}$ \\ \hline
        $g(t=2\text{a})$  & 5.5369 & 5.5348 & 5.5327 & 5.5306 & 5.5286             \\
        $g(t=4\text{a})$  & 6.1175 & 6.1151 & 6.1128 & 6.1104 & 6.1081             \\
        $g(t=6\text{a})$  & 6.4437 & 6.4411 & 6.4386 & 6.4361 & 6.4336             \\
        $g(t=8\text{a})$  & 6.6661 & 6.6634 & 6.6607 & 6.6581 & 6.6555             \\
        $g(t=10\text{a})$ & 6.8322 & 6.8294 & 6.8267 & 6.8240 & 6.8213             \\
        $g(t=12\text{a})$ & 6.9632 & 6.9603 & 6.9576 & 6.9548 & 6.9520             \\
        $g(t=14\text{a})$ & 7.0704 & 7.0675 & 7.0646 & 7.0618 & 7.0590             \\
        $g(t=16\text{a})$ & 7.1603 & 7.1574 & 7.1545 & 7.1516 & 7.1487             \\
        $g(t=18\text{a})$ & 7.2373 & 7.2343 & 7.2314 & 7.2284 & 7.2255             \\
        $g(t=20\text{a})$ & 7.3042 & 7.3011 & 7.2982 & 7.2952 & 7.2923             \\ \hline
    \end{tabular}
\end{table}

\begin{table}[]
    \centering
    \caption{$g$-function of 2 boreholes given by optimal nonuniform discretization with regularization}
    \label{tab: reg-4}
    \begin{tabular}{|l|l|l|l|l|l|}
        \hline
        $N_{s}$           & 64     & 128    & 256                & 512                & 1024               \\ \hline
        Condition number  & 765.6  & 16357  & $1.12\times10^{7}$ & $3.24\times10^{9}$ & $4.57\times10^{9}$ \\ \hline
        $g(t=2\text{a})$  & 5.5331 & 5.5310 & 5.5290             & 5.5276             & 5.5273             \\
        $g(t=4\text{a})$  & 6.1132 & 6.1109 & 6.1086             & 6.1070             & 6.1067             \\
        $g(t=6\text{a})$  & 6.4391 & 6.4366 & 6.4341             & 6.4325             & 6.4321             \\
        $g(t=8\text{a})$  & 6.6612 & 6.6586 & 6.6560             & 6.6543             & 6.6539             \\
        $g(t=10\text{a})$ & 6.8272 & 6.8245 & 6.8218             & 6.8200             & 6.8196             \\
        $g(t=12\text{a})$ & 6.9581 & 6.9553 & 6.9526             & 6.9507             & 6.9503             \\
        $g(t=14\text{a})$ & 7.0652 & 7.0623 & 7.0595             & 7.0576             & 7.0572             \\
        $g(t=16\text{a})$ & 7.1550 & 7.1521 & 7.1493             & 7.1474             & 7.1469             \\
        $g(t=18\text{a})$ & 7.2319 & 7.2290 & 7.2261             & 7.2242             & 7.2237             \\
        $g(t=20\text{a})$ & 7.2987 & 7.2958 & 7.2928             & 7.2909             & 7.2904             \\ \hline
    \end{tabular}
\end{table}

\subsection{Improvement on the computational efficiency and accuracy}
The 2 borehole test case above has proved the necessity of regularization and the fast convergence rate of the proposed discretization to obtain accurate values of $g$-function compared with existing uniform and nonuniform discretization methods based on SFLS model. It also demonstrates the convergence behavior of $g$-function obtained by proposed discretization scheme as the number of quadrature points for heat $N_{quad}^{heat}$ increases with fixed number of quadrature points for temperature in each interval $N_{quad}^{temp}=40$. Now we further investigate how the proposed method improves the computational efficiency and accuracy of $g$-function compared with existing SFLS models.

Firstly, we analyze how the discretization to evaluate average temperature $N_{quad}^{temp}$ influences the accuracy of $g$-function and how to select the appropriate combination of $N_{quad}^{heat}$ and $N_{quad}^{temp}$ to achieve the acceptable accuracy with the least computational cost.

We choose the same two boreholes as the test case and calculate the $g$-function at ten time points from 2 years to 20 years with the combination between $N_{quad}^{heat}=[64, 128, 256, 512]$ and $N_{quad}^{temp}=[2, 3, 5, 10, 25, 50, 75]$. The accuracy is evaluated by the relative error between the approximate value and the assumed exact value of $g$-function when $N_{quad}^{heat}=1024$ and $N_{quad}^{temp}=50$. The regularization parameter $\lambda$ is set to $1\times 10^{-8}$.

\begin{equation}
    Error=\frac{\displaystyle\sum_{k=1}^{10}g_{approx}(t_k) - \displaystyle\sum_{k=1}^{10}g_{exact}(t_k)}{\displaystyle\sum_{k=1}^{10}g_{exact}(t_k)}
\end{equation}

\begin{table}[]
    \centering
    \caption{Relative error of $g$-function given by the proposed discretization scheme with different combinations of $N_{quad}^{heat}$ and $N_{quad}^{temp}$ in 2 borehole test case}
    \label{tab: 2borehole_combination}
    \begin{tabular}{c|c|c|c|c}
        \diagbox{{$N_{quad}^{temp}$}}{{$N_{quad}^{heat}$}} & 64        & 128       & 256       & 512       \\ \hline
        2                                                  & -0.194141 & -0.107730 & -0.041976 & -0.008899 \\
        3                                                  & 0.339433  & 0.112407  & 0.024122  & 0.002487  \\
        5                                                  & 0.165572  & 0.043313  & 0.005849  & 0.000239  \\
        10                                                 & -0.033025 & -0.004897 & 0.000151  & 0.000028  \\
        25                                                 & 0.003759  & 0.000929  & 0.000366  & 0.000029  \\
        50                                                 & 0.001378  & 0.000881  & 0.000366  & 0.000029  \\
        75                                                 & 0.001414  & 0.000881  & 0.000366  & 0.000029  \\
    \end{tabular}
\end{table}

The results are shown in Table \ref{tab: 2borehole_combination}. We can observe that the relative error of $g$-function generally decreases as the number of quadrature points for heat $N_{quad}^{heat}$ increases, which is consistent with the convergence rate we discussed before. When $N_{quad}^{heat}$ is fixed, the relative error converges to a stable value with oscillations as the number of quadrature points to evaluate average temperature $N_{quad}^{temp}$ in each interval increases. It is notable that the convergence rate with respect to $N_{quad}^{temp}$ is much faster with bigger $N_{quad}^{heat}$. For example, the relative error of $g$-function with $N_{quad}^{heat}=128$ converges to 0.000881 when $N_{quad}^{temp}=50$, while the relative error with $N_{quad}^{heat}=512$ converges to 0.000028 with only $N_{quad}^{temp}=10$. This phenomenon is intuitive because when the interval to evaluate average temperature shrinks as $N_{quad}^{heat}$ increases, its temperature distribution becomes more flat and we need fewer quadrature points $N_{quad}^{temp}$ to approximate the average temperature in each interval.

The observation indicates that the value of $N_{quad}^{temp}$ should be selected appropriately according to the order of discretization $N_{quad}^{heat}$. From the perspective of computational complexity, constructing the linear system \ref{eq: Linear system} needs to compute the impact of $N_b N_{quad}^{heat}$ point heat sources on the temperature of $N_b N_{quad}^{heat} N_{quad}^{temp}$ quadrature points, which has a $O(N_b^2 {N_{quad}^{heat}}^2 N_{quad}^{temp})$ complexity proportional with $N_{quad}^{temp}$. Meanwhile, increasing $N_{quad}^{temp}$ beyond the converging threshold indefinitely does not improve the accuracy of $g$-function significantly. Therefore, we suggest to set bigger $N_{quad}^{temp}$ when $N_{quad}^{heat}$ is small, and decrease $N_{quad}^{temp}$ to a smaller value like 5 or 10 when $N_{quad}^{heat}$ is sufficiently large to achieve a good balance between accuracy and computational cost.

Based on the analysis above, we choose the following 4 configurations of discretization to compare both the computational efficiency and accuracy of the proposed discretization scheme against existing methods in the 2 borehole test case as shown in Table \ref{tab: 2borehole_efficiency_settings}.

\begin{table}[]
    \centering
    \caption{Configurations of discretization to compare the computational efficiency and accuracy}
    \label{tab: 2borehole_efficiency_settings}
    \begin{tabular}{c|c|c}
        No & Proposed method ($N_{quad}^{heat}$, $N_{quad}^{temp}$) & Other SFLS models ($N_s$) \\ \hline
        1  & (64, 50)                                               & 64                        \\
        2  & (128, 50)                                              & 128                       \\
        3  & (256, 25)                                              & 256                       \\
        4  & (512, 10)                                              & 512                       \\
    \end{tabular}
\end{table}

\begin{table}[]
    \centering
    \caption{Comparison of the computational efficiency and accuracy in 2 borehole test case}
    \label{tab: 2borehole_efficiency_result}
    \begin{tabular}{c|cc|cc|cc|cc}
        \multicolumn{1}{c|}{\multirow{2}{*}{No}} &
        \multicolumn{2}{c|}{Proposed}            &
        \multicolumn{2}{c|}{Uniform}             &
        \multicolumn{2}{c|}{Nonuniform-base}     &
        \multicolumn{2}{c}{Nonuniform-opt}
        \\ \cline{2-9}
        \multicolumn{1}{c|}{}                    &
        Time(s)                                  &
        Error                                    &
        Time(s)                                  &
        Error                                    &
        Time(s)                                  &
        Error                                    &
        Time(s)                                  &
        Error
        \\ \hline
        1                                        & \textbf{0.09} & \num{1.4e-3}          & 1.96  & \num{3.5e-3} & 1.80   & \num{2.0e-3} & 1.69   & \textbf{\num{1.3e-3}} \\
        2                                        & \textbf{0.41} & \textbf{\num{8.8e-4}} & 10.71 & \num{2.8e-3} & 12.00  & \num{1.6e-3} & 10.70  & \num{9.2e-4}          \\
        3                                        & \textbf{0.83} & \textbf{\num{3.7e-4}} & 46.89 & \num{2.3e-3} & 48.27  & \num{1.2e-3} & 52.54  & \num{5.2e-4}          \\
        4                                        & \textbf{1.36} & \textbf{\num{2.8e-5}} & 190.1 & \num{1.9e-3} & 203.21 & \num{8.4e-4} & 222.78 & \num{2.6e-4}          \\
    \end{tabular}
\end{table}

The total computational time to evaluate the $g$-function from 2 years to 20 years and their relative errors given by four discretimzation methods are shown in Table \ref{tab: 2borehole_efficiency_result} with the following observations:

\begin{enumerate}
    \item Three SFLS models with different discretization schemes exhibit similar computational efficiency with nearly quadratic complexity with respect to the order of discretization, because they all need to compare the $N_b^2 N_s^2$ segment-to-segment response factors between $N_b N_s$ segments. The computational time with 512 segments per borehole has exceeded 200 seconds even for the 2 borehole system. In contrast, the proposed discretization scheme is significantly faster with acceleration rate from 20 times to 160 times. The acceleration can be explained from two aspects: firstly the point-to-point heat response factor in our method is analytical and more efficient to compute than the segment-to-segment response factor containing double integral in SFLS models; secondly, the adaptive selection of discretization parameter ($N_{quad}^{heat}$, $N_{quad}^{temp}$) mitigates the $O(N_b^2 {N_{quad}^{heat}}^2 N_{quad}^{temp})$ complexity by reducing $N_{quad}^{temp}$ for bigger $N_{quad}^{heat}$, thus the computational time grows slower than quadratic with respect to $N_{quad}^{heat}$ and achieves higher acceleration rate under fine-grained discretization (160 times acceleration when $N_{quad}^{heat}=512$ compared to 20 times acceleration when $N_{quad}^{heat}=64$).
    \item Besides the computational efficiency, the proposed discretization scheme also achieves better accuracy than existing methods especially in the context of fine-grained discretization (the only exception being the case with $N_{quad}^{heat}=64$ but the relative error is close to the optimal nonuniform discretization for SFLS model). The relative error of $g$-function given by the proposed method is less than $2.8\times 10^{-5}$ with $N_{quad}^{heat}=512$, while the relative error of SFLS models can only reach $2.6\times 10^{-4}$ even with the optimal nonuniform discretization. The proposed method achieves nearly 10 times improvement on accuracy compared to the best discretization scheme of SFLS models.
\end{enumerate}

We also apply the proposed method to larger test cases with more boreholes including a 2x2 and 3x3 borehole configuration to demonstrate its accuracy and efficiency. We reduce the timespan to 10 years (5 timepoints from 2 years to 10 years) to decrease the computational cost and only compare the proposed method with the optimal nonuniform discretization for SFLS model. The results are shown in Tables \ref{tab: 2x2case} and \ref{tab: 3x3case} and we can observe that the proposed method still achieves approximately 20-150 times better computational efficiency with close or better accuracy than the optimal nonuniform discretization for SFLS model.

\begin{table}[]
    \centering
    \caption{Comparison of the computational efficiency and accuracy in $2 \times 2$ borehole test case}
    \label{tab: 2x2case}
    \begin{tabular}{c|cc|cc}
        \multicolumn{1}{c|}{\multirow{2}{*}{No}} &
        \multicolumn{2}{c|}{Proposed}            &
        \multicolumn{2}{c}{Nonuniform-opt}
        \\ \cline{2-5}
        \multicolumn{1}{c|}{}                    &
        Time(s)                                  &
        Error                                    &
        Time(s)                                  &
        Error
        \\ \hline
        1                                        & \textbf{0.21} & \num{1.5e-3}          & 4.57   & \textbf{\num{1.4e-3}} \\
        2                                        & \textbf{0.72} & \textbf{\num{9.1e-4}} & 22.62  & \num{9.5e-4}          \\
        3                                        & \textbf{1.56} & \textbf{\num{3.8e-4}} & 97.91  & \num{5.4e-4}          \\
        4                                        & \textbf{2.67} & \textbf{\num{2.9e-5}} & 400.82 & \num{2.7e-4}          \\
    \end{tabular}
\end{table}

\begin{table}[]
    \centering
    \caption{Comparison of the computational efficiency and accuracy in $3 \times 3$ borehole test case}
    \label{tab: 3x3case}
    \begin{tabular}{c|cc|cc}
        \multicolumn{1}{c|}{\multirow{2}{*}{No}} &
        \multicolumn{2}{c|}{Proposed}            &
        \multicolumn{2}{c}{Nonuniform-opt}
        \\ \cline{2-5}
        \multicolumn{1}{c|}{}                    &
        Time(s)                                  &
        Error                                    &
        Time(s)                                  &
        Error
        \\ \hline
        1                                        & \textbf{0.96} & \num{1.6e-3}           & 22.63   & \textbf{\num{1.5e-3}} \\
        2                                        & \textbf{3.84} & \textbf{\num{9.96e-4}} & 106.43  & \num{1.03e-3}         \\
        3                                        & \textbf{7.89} & \textbf{\num{4.1e-4}}  & 447.27  & \num{5.8e-4}          \\
        4                                        & \textbf{14.3} & \textbf{\num{3.2e-5}}  & 1874.84 & \num{2.9e-4}          \\
    \end{tabular}
\end{table}

\section{Conclusions}
\label{section: conclusion}

This paper provides a novel discretization method for calculating $g$-functions of vertical geothermal boreholes, based on the direct numerical solution of the spatio-temporal integral equations that describe the thermal process. This method is significantly superior to the traditional SFLS model in both computational speed and accuracy, especially when using high-order discretization. Its core advantage lies in replacing the time-consuming segment-to-segment thermal response factors in the SFLS model, which require numerical integration, with point-to-point response factors that have an analytical form and are more efficient to compute. This methodological shift means that the $g$-function calculation no longer relies on empirical segmentation strategies but is instead theoretically supported by the Gauss-Legendre quadrature rule.

Furthermore, this paper identifies the $g$-function calculation problem as a Fredholm integral equation of the first kind and reveals the root cause for the divergence of calculation results at high discretization orders: the severe ill-conditioning of the discretized linear system. To solve this, we introduced a regularization technique that effectively overcomes this numerical instability, ensuring that the $g$-function values converge stably at any order of discretization.

The proposed method is valuable for applications such as optimization-based algorithms for borehole field design, which require a large number of simulations. Numerical tests clearly demonstrate its performance advantages: when processing a 3x3 borehole field, the proposed method reduces the computation time from 1874.84 seconds to just 14.3 seconds compared to the state-of-the-art nonuniform SFLS method, while simultaneously improving accuracy by nearly an order of magnitude.

\bibliographystyle{elsarticle-num}
\bibliography{main.bib}

\end{document}